\documentclass[prd,nofootinbib,superscriptaddress]{revtex4}

\usepackage{amsmath}
\usepackage{graphicx}
\usepackage{color}

\usepackage{amsmath}
\usepackage{graphicx}

\newcommand{\beq}{\begin{equation}}
\newcommand{\eeq}{\end{equation}}

\begin{document}

\title{Magnetic Field Probe of the No-Hair Theorem}
\author{Pierre Christian\\
Astronomy Department, Harvard University, 60 Garden St., Cambridge, MA 02138}
\email{pchristian@cfa.harvard.edu}

\begin{abstract}
We discuss the consequences of violating the no-hair theorem on magnetic fields surrounding a black hole. This is achieved by parametrically deforming the Kerr spacetime and studying the effects of such deformations on asymptotically uniform magnetic fields around the black hole. We compute the deformed electromagnetic field for slow spins and small deformation parameter, and show that the correction is of order the deformation parameter and mimics the angular structure of a quadrupole. 
\end{abstract}

\maketitle

\section{Introduction}
The no-hair theorem of General Relativity (GR) states that isolated, stationary black holes are described by only three parameters: $M$, the mass of the black hole; $J$, the spin of the black hole; and $Q$, the charge of the black hole \citep{NoHair1, NoHair2, NoHair3, NoHair4, NoHair5}. In terms of the metric, this means that the most general black holes satisfying the no-hair theorem is the Kerr-Newman metric describing a charged, rotating black hole. However, in typical astrophysical settings, charge neutrality is expected. This reduces the metric to the Kerr metric describing an uncharged, rotating black hole, given in Boyer-Lindquist coordinates $(t, r ,\theta, \phi)$ as

\begin{align}
ds^2 &= - \left(1 - \frac{2 M r} {\Sigma} \right) dt^2 - \frac{4 M a r \sin^2 \theta}{\Sigma} dt d\phi + \frac{\Sigma}{\Delta} dr^2  + \Sigma d\theta^2 + \left(r^2 + a^2 + \frac{2 M a^2 r \sin^2 \theta}{\Sigma} \right) \sin^2 \theta d\phi^2\;,
\end{align}
where $G=c=1$, $a = J/M$, $\Sigma=r^2 + a^2 \cos^2 \theta$, $\Delta = r^2 - 2 Mr + a^2$, and we have taken the $(-,+,+,+)$ as the metric signature. 

In the derivation of the no-hair theorem, it is necessary to assume that the black hole spacetime does not possess either naked singularities or closed timelike curves outside of a horizon. Therefore, the detection of a black hole which violates the no-hair theorem implies that either GR, the Cosmic Censorship Conjecture, or the Chronologic Censorship Conjecture is invalid \citep{JSysStudy}. 

There has been multiple proposals in the past for testing the black hole no-hair theorem, with most of them focusing on methods to observe the black hole quadrupole. For an uncharged black hole, the no-hair theorem demands that all multipole moments of the black hole depend only on $M$ and $J$. In particular, the dimensionless black hole quadrupole, $q$ is given by
\beq \label{eq:Quadrupole}
q \equiv \frac{c^4 Q}{G^2 M^3} = - \left[ \frac{c J}{G M^2} \right]^2 \; ,
\eeq
where $Q$ is the black hole's quadrupole moment, and we have reintroduced the factors of $c$ and $G$. Violations of equation (\ref{eq:Quadrupole}) causes astrophysical observables like the relativistically broadened iron lines \citep{JPtestIronLines}, the shape of the black hole shadow \citep{JPtest1}, and the Shapiro delay to be modified from their Kerr counterparts \citep{ChristianPulsar}. 

In this work we propose that the magnetic field structure around the black hole will also be modified by the presence of a non-Kerr quadrupole. This change could in principle be detected by observational campaigns designed to probe magnetic fields close to the black hole horizon \citep{Gold}. Furthermore, this calculation is important for testing force-free numerical computations. 

In particular, we are interested in black holes immersed in an asymptotically uniform external magnetic field which shares the same symmetries of the spacetime. The magnetic field is considered to be a test field that does not affect the spacetime geometry, and is assumed to satisfy the source-free Maxwell's equations. Examples where such a condition is realized in an astrophysical setting is when a magnetar orbits a black hole within its light cylinder \citep{DOrazio1, DOrazio2} or for a black hole immersed in tenuous plasma \citep{MovingWald}.  

The main machinery of this work is a theorem by Wald \citep{Wald} which states that in GR the behavior of electromagnetic test fields around an asymptotically flat, axisymmetric, vacuum spacetimes is related to the spacetime Killing vectors. While this theorem has been extended to a variety of non-GR gravitational theories \citep{VacuumWald,NonGR1, NonGR2}, we will only need the GR version here. The reason for this is twofold: first, we want to be agnostic towards the particular theoretical extension of GR; and second, it is possible that even within GR the no-hair theorem is violated \citep{MNmetric}. To this end we employ a metric that parametrically deforms the Kerr spacetime, and compute the effects of the deformation parameter on the test magnetic field. This approach was first attempted by \cite{Ahmedov1} for the Johannsen-Psaltis (JP) metric \citep{JPMetric}. However, the JP metric is not Ricci flat \citep{JSysStudy, JPMetric}, rendering the Wald solution invalid.

The organization of this article is as follows: in \S2 we discuss the quasi-Kerr (QK) metric, a parametric deformation of the Kerr metric that we employ in our calculation; in \S3 we review the Wald solution; in \S4 we compute the Wald solution for the QK metric in the Boyer-Lindquist like coordinates; in \S5 we transform the solution to the Zero Angular Momentum Observer (ZAMO) frame; and in \S6 we provide some concluding remarks.

\section{The quasi-Kerr metric}
The QK metric \citep{quasiKerr} is a parametric deformation of the Kerr metric given by
\beq \label{eq:metricmodif}
g_{\mu \nu} = g_{\mu \nu}^{\rm Kerr} + \epsilon h_{\mu \nu} \; ,
\eeq
where $ g_{\mu \nu}^{\rm Kerr}$ is the Kerr metric, $\epsilon$ a small parameter, and $h_{\mu \nu}$ is given by
\begin{align}
h^{tt} &= (1- 2M/r)^{-1} \left[ (1-3 \cos^2 \theta) \mathcal{F}_1(r) \right] \nonumber  \;, \\
h^{rr} &= (1- 2M/r) \left[ (1-3 \cos^2 \theta) \mathcal{F}_1(r) \right] \nonumber \;,  \\
h^{\theta \theta} &= -r^{-2} \left[ (1-3 \cos^2 \theta) \mathcal{F}_2(r) \right] \nonumber \;,  \\
h^{\phi \phi} &= -(r \sin \theta)^{-2} \left[ (1-3 \cos^2 \theta) \mathcal{F}_2(r) \right]  \;,
\end{align} 
where the coordinates $(t, r ,\theta, \phi)$ are Boyer-Lindquiest like, and the functions $\mathcal{F}_1(r)$ and $\mathcal{F}_2(r)$ are given in Appendix A of \cite{quasiKerr} as
\begin{align}
\mathcal{F}_1(r) = &- \frac{5(r-M)}{8 Mr (r-2M)} (2 M^2 +6 Mr - 3r^2) - \frac{15 r (r-2M)}{16 M^2} \ln \left( \frac{r}{r-2M} \right) \; ,
\end{align}
\begin{align}
\mathcal{F}_2(r) = &\frac{5}{8 Mr} (2 M^2 - 3 Mr - 3r^2) + \frac{15}{16M^2} (r^2 - 2 M^2) \ln \left( \frac{r}{r-2M} \right) \; .
\end{align}
The $\epsilon$ parameter of the QK metric modifies the quadrupole moment, Q, of the black hole into \citep{JSysStudy}
\beq
Q = - M(a^2 + \epsilon M^2) \; ,
\eeq 
where the $a^2$ piece is the quadrupole moment of the standard Kerr black hole. 

The QK metric is stationary and axisymmetric, admitting a timelike Killing vector $\eta^\nu$ and an axisymmetric Killing vector $\psi^\nu$. Furthermore, it is asymptotically flat and satisfy the vacuum Einstein equation for low spins and small $\epsilon$. Indeed, neglecting terms of order $O(a^2)$, $O(\epsilon a)$, and $O(\epsilon^2)$, the metric is Ricci flat \citep{JSysStudy}. In this article we will work exclusively in these regimes. 

\section{Wald magnetic field solution}
If we immerse a black hole in an external magnetic field, the immense curvature of the spacetime modifies the magnetic field close to the black hole. If the magnetic field is a test field (i.e. small enough to not disturb the spacetime itself), it is required to satisfy the source-free Maxwell's equations,
\beq
\nabla_\nu F^{\mu \nu} = 0 \; ,
\eeq
where $F^{\mu \nu}$ is the electromagnetic tensor. Wald found that for a magnetic field that is asymptotically parallel to the rotation axis of the black hole \citep{Wald},
\begin{align}
\vec{B} &= B_0 \hat{z} \; ,
\end{align}
at spatial infinity, where $\hat{z}$ is the direction parallel to the black hole's rotation axis, the solution of the source-free Maxwell's equations is given by
\beq
F = \frac{1}{2} B_0 \left( d\psi + 2 a  d\eta \right) \; ,
\eeq 
where $a$ is the spin of the black hole, while $d\psi$ and $d\eta$ refers to the one-forms corresponding to the Killing vectors $\psi^\nu$ and $\eta^\nu$, defined by
\begin{align}
\eta &\equiv \eta_\nu dx^\nu \;, \\
\psi &\equiv \psi_\nu dx^\nu \;. 
\end{align}
This solution is valid as long as the spacetime satisfies the vacuum Einstein equation (Ricci flat),
\beq
R_{\mu \nu} = 0 \; .
\eeq
As the QK metric is Ricci flat when the spin and deformation parameter are small, we can use this method to solve the source-free Maxwell's equations for QK black holes in these regimes. 

\section{QK black hole immersed in magnetic field}

In the coordinates we are using, the Killing vectors of the QK black hole is identical to the usual Kerr Killing vectors in Boyer-Lindquist coordinates,
\begin{align}
\eta^\nu &= \frac{\partial}{\partial t} \; ,\\
\psi^\nu &= \frac{\partial}{\partial \phi} \; .
\end{align}
Therefore, we can rewrite the Wald solution in terms of the metric components via the identifications
\begin{align}
\eta_\nu &= g_{\nu t} \; ,\\
\psi_\nu &= g_{\nu \phi} \; .
\end{align}
In particular, due to its dependence on $d\psi$ and $d\eta$, the electromagnetic tensor $F$ will consist of terms proportional to derivatives of the metric $\partial_\alpha g_{\mu \nu}$. 

Churning through these derivatives, we obtain the following components of $F$ up to second order in the spin parameter $a$: 
\beq
F_{tr} = - B_0 \frac{4 a M (\sin^2 \theta -1) }{r^2} \; .
\eeq
\beq
F_{t\theta} = B_0 \frac{8 a M \cos \theta \sin \theta }{r} \; ,
\eeq
\begin{align}
F_{r\phi} &= 2 B_0 r \sin ^2\theta + \epsilon B_0 \frac{5 (M+r)  \left[1+3 \cos (2 \theta )\right] \left[2 M \left(M^2-6 M r+3 r^2\right)-3 r \left(2 M^2-3 M
   r+r^2\right) \log \left(\frac{r}{-2 M+r}\right)\right] \sin ^2(\theta )}{8 M^2 (2 M-r)}
\end{align}
\begin{align}
F_{\theta \phi} &= 2 B_0 r^2 \sin\theta \cos \theta + \epsilon B_0 \frac{5 r   \left[ 2 M \left(2 M^2-3 M r-3 r^2\right)+3 r \left(-2 M^2+r^2\right) \log
   \left(\frac{r}{-2 M+r}\right)\right] \left[2 \sin (2 \theta )-3 \sin (4 \theta )\right]}{32 M^2}
   \end{align}
   
\begin{figure}[h!]
\centering
\includegraphics[width=5in]{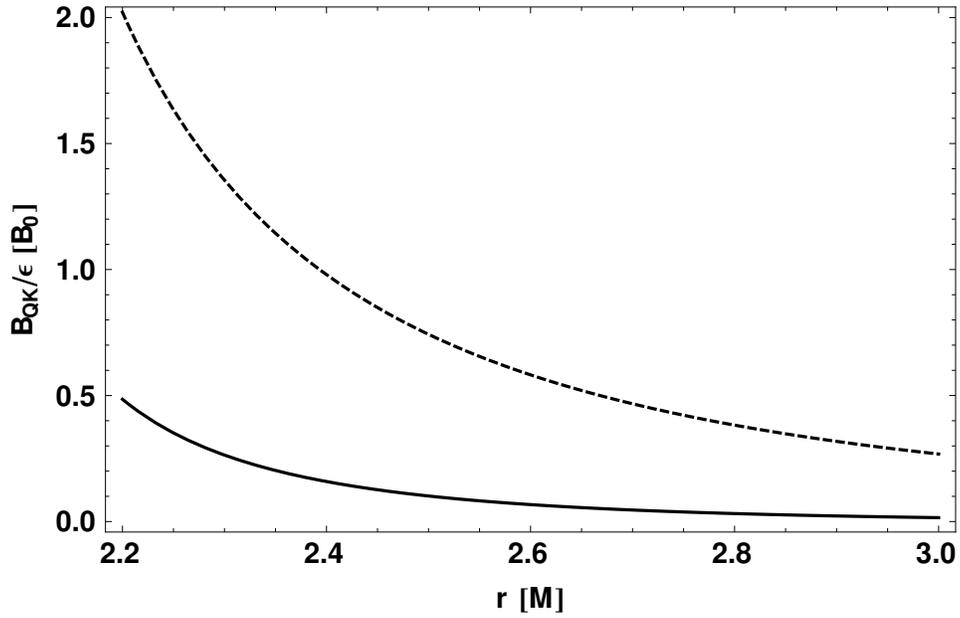}
\caption{$B^{\hat{\theta}_{QK}}$ (solid) and $B^{\hat{r}_{QK}}$ (dashed) as a function of radius from the black hole for $\theta=\pi/4$. Close to the black hole, the corrections due to the quadrupole modification is of order $\epsilon$. }
\label{fig:Bsolo}
\end{figure}

\begin{figure}[h!]
\centering
\includegraphics[width=5in]{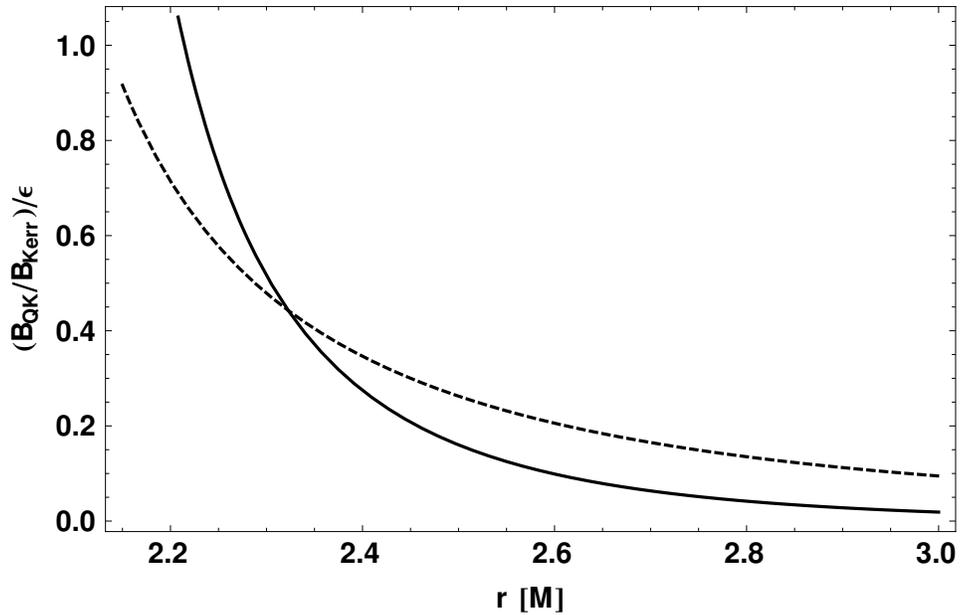}
\caption{The ratio of $B^{\hat{\theta}}_{QK}$ (solid) and $B^{\hat{r}}_{QK}$ (dashed) to the Kerr solution as a function of radius from the black hole for $\theta = \pi/4$. Close to the black hole, the corrections due to the quadrupole modification can exceed that of the Kerr contribution.}
\label{fig:BoverKerr}
\end{figure}

Note that to order $a^2$, the electric field is identical to the Kerr solution. This is because the electric field is generated by the frame dragging of the magnetic field by the rotation of the black hole. Therefore, the electric field terms are at least of order $a$. Higher order corrections to the electric field due to violations of the no-hair theorem is neglected in our approximation. As a result, the Wald charge accumulated by a slowly spinning QK black hole is identical to that of a Kerr black hole. 

\section{Fields in the ZAMO Frame}
In order to obtain the electric and magnetic fields from $F$, a frame must be specified. To this end, we specify the Zero Angular Momentum Observer (ZAMO) frame of the QK metric up to order $\epsilon$:

\begin{align}
e^{\hat{t}}_{\;t} &= \sqrt{1-\frac{2 M}{r}}+\epsilon \frac{5  \left[-1+3 \cos ^2(\theta )\right] \left[-2 M (M-r) \left(2 M^2+6 M r-3
   r^2\right)+3 r^2 (-2 M+r)^2 \log \left(\frac{r}{-2 M+r}\right)\right]}{32 M^2 \sqrt{1-\frac{2 M}{r}} r^2} \; ,
\end{align}

\begin{align}
e^{\hat{r}}_{\; r} &= \sqrt{\frac{r}{-2 M+r}} + \epsilon \frac{5  \left[-1+3 \cos ^2(\theta )\right] \left[2 M (M-r) \left[2 M^2+6 M r-3 r^2\right)-3 r^2 (-2 M+r)^2
   \log \left(\frac{r}{-2 M+r}\right)\right]}{32 M^2 (-2 M+r)^2 \sqrt{1+\frac{2 M}{-2 M+r}}} \; ,
\end{align}

\begin{align}
e^{\hat{\theta}}_{\; \theta} &= r - \epsilon \frac{5 \left[-1+3 \cos ^2(\theta )\right] \left[2 M \left(2 M^2-3 M r-3 r^2\right)+3 r \left(-2
   M^2+r^2\right) \log \left(\frac{r}{-2 M+r}\right)\right]}{32 M^2} \; ,
\end{align}

\begin{align}
e^{\hat{\phi}}_{\; t} &= \frac{2 a M \sin^2 \theta}{r^2} - \epsilon \frac{5 a \left[-1+3 \cos ^2(\theta )\right] \left[2 M \left(2 M^2-3 M r-3 r^2\right)+3 r \left(-2 M^2+r^2\right)
   \log \left(\frac{r}{-2 M+r}\right)\right]  \sin ^3\theta }{16 M r^3} \;.
\end{align}

\begin{align}
e^{\hat{\phi}}_{\; \phi} &= r \sin \theta - \epsilon \frac{5 \left[-1+3 \cos ^2(\theta )\right] \left[2 M \left(2 M^2-3 M r-3 r^2\right)+3 r \left(-2 M^2+r^2\right)
   \log \left(\frac{r}{-2 M+r}\right)\right] \sin\theta } {32 M^2} \;,
\end{align}
where the hatted coordinates are that of the ZAMO frame, and all other components of $e^{\hat{\alpha}}_{\; \beta}$ is zero. 

Projecting $F_{\mu \nu}$ to the ZAMO frame, we obtain the following 
\beq
F_{\hat{\mu} \hat{\nu}} = F_{\hat{\mu} \hat{\nu}}^{\rm Kerr} + \epsilon F_{\hat{\mu} \hat{\nu}}^{QK} \; ,
\eeq
where the quasi-Kerr components $F_{\hat{\mu} \hat{\nu}}^{QK}$ are given by

\begin{align}
F_{\hat{r} \hat{\phi}}^{QK}&= -B^{\hat{\theta}}_{QK} = -\frac{5 B_0 \sqrt{\frac{1}{-2 M+r}} \left[1+3 \cos (2 \theta )\right] \sin \theta}{16 M^2 r^{3/2}} \left[2 M (M-3 r) \left(5 M^2+3 M r-3 r^2\right) \right. \nonumber \\ 
&\;\;\;\;\;\;\;\;\;\;\;\;\;\;\;\;\;\;\;\;\;\;\;\;\;\;\;\;\;\;\;\;\;\;\;\;\;\;\;\;\;\;\;\;\;\;\;\;\;\;\;\;\;\;\;\;\;\;\;\;\;\;\;\;\;\;\;\;\;\;  \left. +3 r \left(-6 M^3+M^2 r+7 M r^2-3 r^3\right) \log \left(\frac{r}{-2
   M+r}\right)\right]      \;,
\\
F_{\hat{\theta} \hat{\phi}}^{QK} &= B^{\hat{r}}_{QK} = \frac{15 B_0 \cos \theta  \sin ^2\theta   \left[2 M \left(2 M^2-3 M r-3 r^2\right)+3 r \left(-2 M^2+r^2\right) \log
   \left(\frac{r}{-2 M+r}\right)\right]}{8 M^2 r}
    \;,
\end{align}
where $B^{\hat{\theta}_{QK}}$ and $B^{\hat{r}}_{QK}$ are the $\hat{\theta}$ and $\hat{r}$ components of the magnetic field three-vector in the ZAMO frame. We plotted these components as a function of distance from the black hole in Figure \ref{fig:Bsolo} for $\theta=\pi/4$. We also plot the ratio between the QK components and the Kerr component in Figure $\ref{fig:BoverKerr}$ to show that there are points close to the black hole where the QK components of the magnetic field become as large as that of the Kerr component. Note that this does not invalidate our approximation of working in the limit where $\epsilon$ is a small parameter, as we do not impose that $B_{QK}$ is small, but rather that $\epsilon h_{\mu\nu}$ is small compared to $g_{\mu \nu}^{\rm Kerr}$.

\begin{figure}[h!]
\centering
\includegraphics[width=5in]{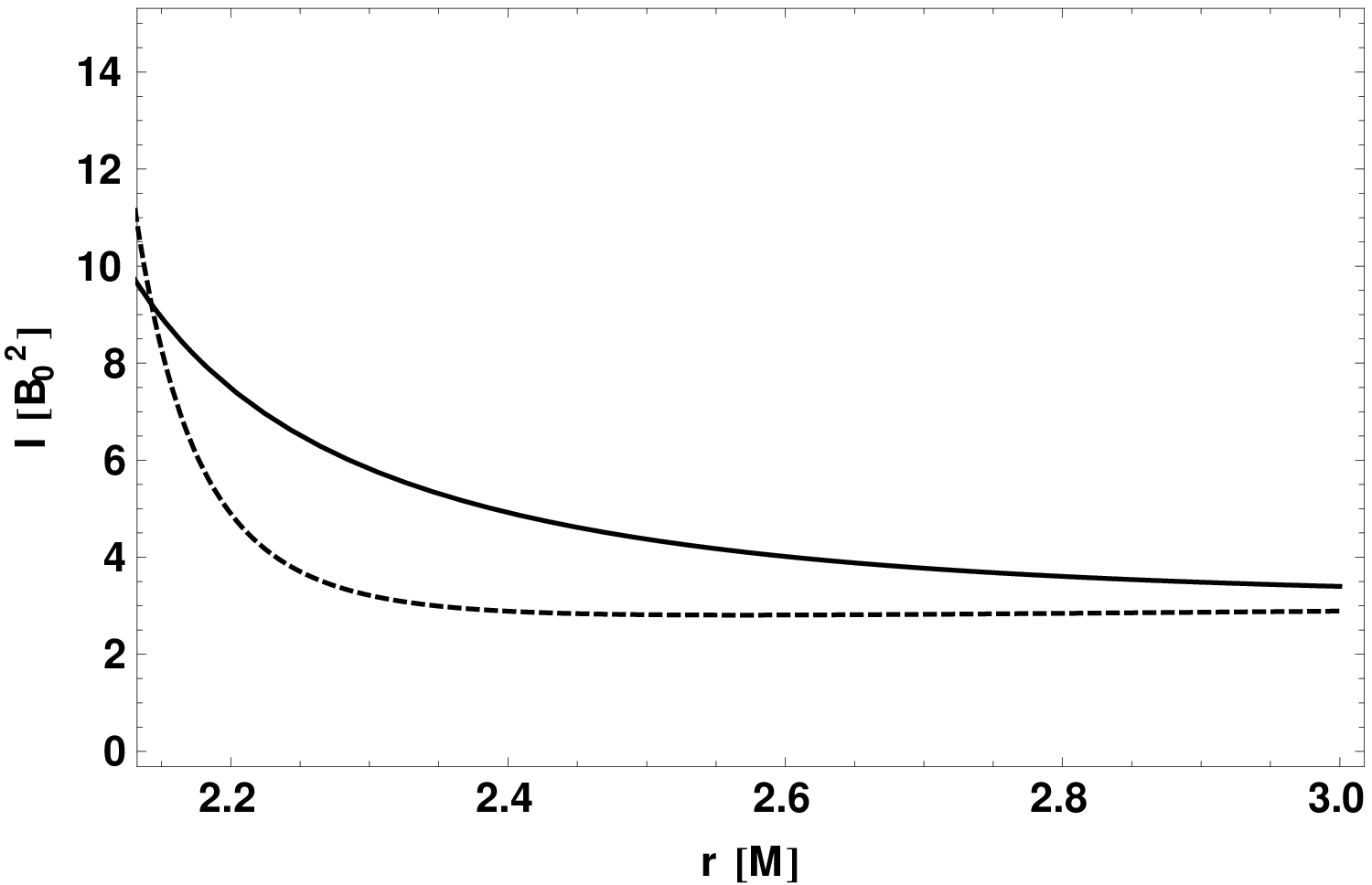}
\includegraphics[width=5in]{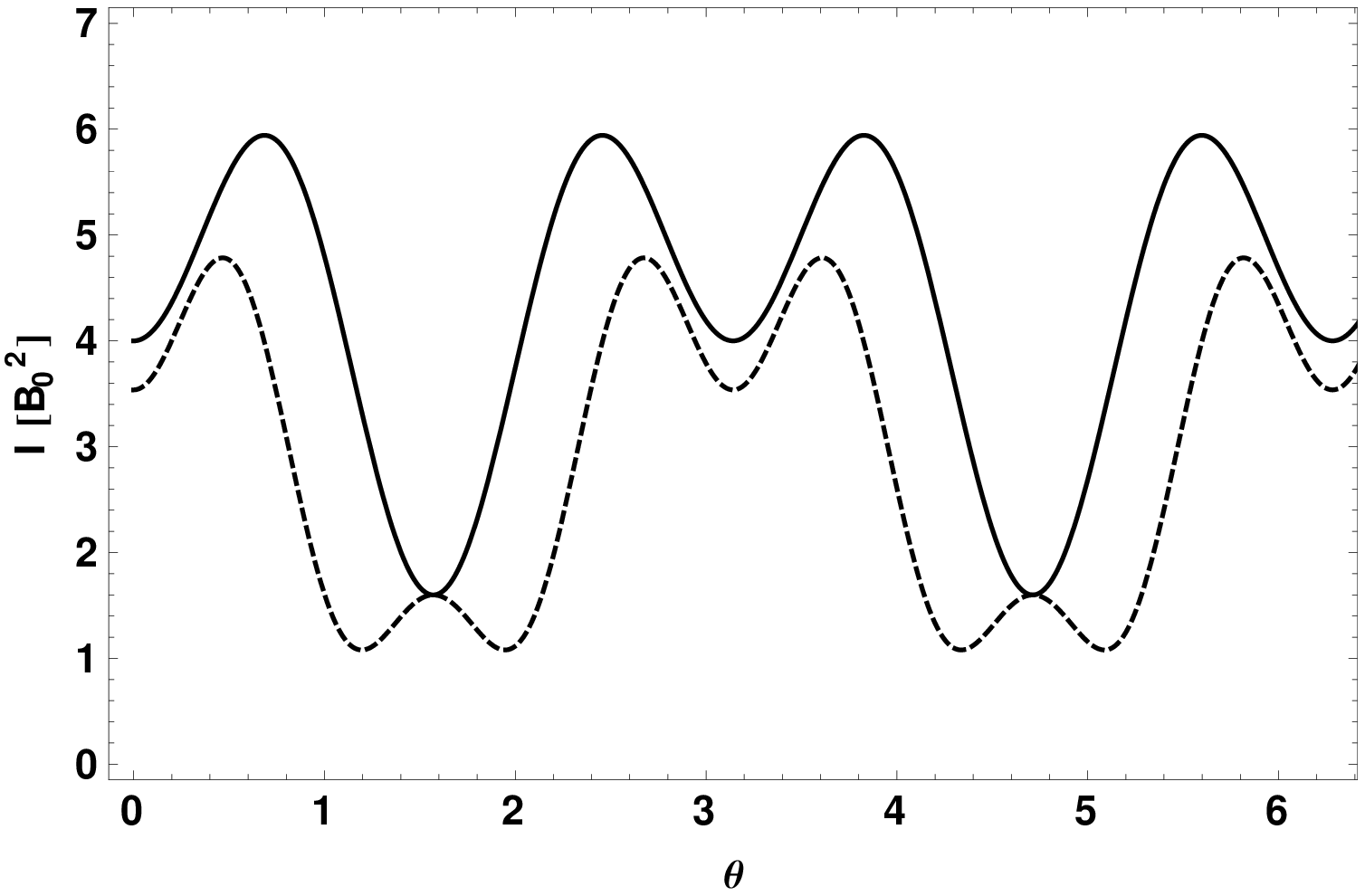}
\caption{The electromagnetic invariant $I=\frac{1}{2} F^{\mu \nu} F_{\mu \nu} = B^2-E^2$ as a function of distance from the black hole and angle for $a=0$ (solid) and $a=0.9$ (dashed). The quadrupolar nature of the electromagnetic field is revealed by the angular structure of $I$.}
\label{fig:minusinvariant}
\end{figure}

In order to present our result in an invariant way, we calculate the electromagnetic invariant 
\beq
 I \equiv \frac{1}{2} F^{\mu \nu} F_{\mu \nu} = B^2-E^2 \; ,
\eeq
of the $\epsilon$ part of the solution as a function of angle and distance from the black hole and plotted them in Figure \ref{fig:minusinvariant}. From the angular structure of $I$, the quadrupolar nature of the electromagnetic field is revealed. 

\section{Conclusion}
We computed the asymptotically uniform magnetic field solution for a black hole that is parametrically deformed from Kerr spacetime using the Wald formalism. We showed that no-hair deformations of the spacetime generates extra fields of strength $\sim \epsilon$ that mimics the quadrupolar structure of the spacetime. Finally, we would like to note that our solution can be transformed to that of an asymptotically uniform electric field by simply taking a Hodge dual of $F_{\mu \nu}$.

\end{document}